\documentclass[runningheads]{llncs}
\usepackage[T1]{fontenc}
\usepackage{graphicx}
\usepackage{subcaption}
\begin{document}
\title{Monte Cimone v2: HPC RISC-V Cluster Evaluation and Optimization}
\author{Emanuele Venieri\inst{1}\orcidID{0009-0007-7182-505X} \and
Simone Manoni\inst{1}\orcidID{0000-0003-1797-1893} \and
Gabriele Ceccolini\inst{1}\orcidID{0009-0004-9402-857X} \and
Giacomo Madella\inst{1}\orcidID{0009-0009-1289-6354} \and
Federico Ficarelli\inst{3}\orcidID{0000-0003-1447-8278} \and
Daniele Gregori\inst{4}\orcidID{0000-0002-6137-6453} \and
Andrea Acquaviva\inst{1}\orcidID{0000-0002-7323-759X} \and
Luca Benini\inst{1,2}\orcidID{0000-0001-8068-3806} \and
Andrea Bartolini\inst{1}\orcidID{0000-0002-1148-2450}}
\authorrunning{E. Venieri et al.}
\institute{University of Bologna, Italy \and ETH Zürich University, Switzerland \and CINECA, Italy \and E4 Computer Engineering Spa, Italy} %\and
%Springer Heidelberg, Tiergartenstr. 17, 69121 Heidelberg, Germany
\maketitle              % typeset the header of the contribution
\begin{abstract}
Many RISC-V (RV) platforms and SoCs have been announced in recent years targeting the HPC sector, but only a few of them are commercially available and engineered to fit the HPC requirements. The Monte Cimone project targeted assessing their capabilities and maturity, aiming to make RISC-V a competitive choice when building a datacenter. Nowadays, Systems-on-chip (SoCs) featuring RV cores with vector extension, form factor and memory capacity suitable for HPC applications are available in the market, but it is unclear how compilers and open-source libraries can take advantage of its performance.  In this paper, we describe the performance assessment of the upgrade of the Monte Cimone (MCv2) cluster with the Sophgo SG2042 processor on HPC workloads. Also adding an exploration of BLAS libraries optimization. The upgrade increases the attained node's performance by 127x on HPL DP FLOP/s and 69x on Stream Memory Bandwidth.

\end{abstract}
\section{Introduction}
In recent years, RISC-V has gained significant traction as an open-standard Instruction Set Architecture (ISA) with its open-source, modular, and extensible design. Early academic and commercial implementation of RISC-V platforms targeted mainly low-power embedded systems and microcontrollers \cite{gap}.
This initial focus served as a natural starting point, as embedded platforms are inherently simpler from an architectural design perspective and allowed for the incremental development of the RISC-V ecosystem, ensuring the gradual maturation of compilers, software toolchains, and system-level optimizations.

In the last few years, advancements in RISC-V hardware and software ecosystems have driven its evolution toward higher-performance platforms. The introduction of 64-bit processors \cite{cva6}, vector extensions (RVV) \cite{ara}, \cite{vitru}, and improved memory subsystems has enabled RISC-V to scale beyond embedded computing, making it increasingly viable for high-performance computing (HPC) workloads.

As the push for RISC-V in HPC gained momentum, early efforts focused on validating the potential of this open ISA for highly demanding computational tasks. As part of this exploration, Monte Cimone (MCv1)\cite{mc} was developed as the first pioneering multi-node computing platform to assess the maturity of RISC-V HPC system software. Designed to address the challenges of integrating multi-node RISC-V clusters, MCv1 serves as a testbed for building a comprehensive HPC stack, including interconnects, storage and power monitoring, all using RISC-V hardware.

In this paper, we propose a major upgrade of Monte Cimone named MCv2 to assess the maturity of software libraries and compilers
in leveraging novel, more compute-capable RISC-V many-core processors featuring vector extensions and larger memory capacity. The key contributions include:
\begin{itemize}
    \item The description of the hardware architecture of the new MCv2 nodes and the hardware-software infrastructure we developed for performance analysis.
    \item The enhancement of the software stack with compilation toolchains and optimized BLAS libraries both from the community and produced in this work.
    \item An extensive benchmarking campaign leveraging HPC-class tools to evaluate performance, efficiency, and scalability, which provides a comprehensive insight into the state-of-the-art in RISC-V HPC architectures.
\end{itemize}

\section{Background}
Monte Cimone is a multi-node compute cluster built on RISC-V architecture, designed as a validation platform for HPC systems.
The first iteration of the Monte Cimone (MCv1) \cite{mc} compute cluster utilized four E4 RV007 Server Blades, based on two boards SiFive HiFive Unmatched featuring the SiFive Freedom U740 SoC. The peak theoretical performance was 4.0 Gflop/s per node.
The Monte Cimone software stack is built with Spack and accessible via shared modules. Nodes run Ubuntu 21.04, and mount a shared NFS, the job scheduler is SLURM and the system monitor is ExaMon\cite{examon}. The system achieves 12.65 Gflop/s for full-machine HPL\cite{hpl} and of 1.1 GB/s for STREAM DDR bandwidth\cite{stream} benchmarks.

Since the first iteration of the Monte Cimone cluster, new and enhanced hardware has become available. Notably, a new processor has emerged in high-performance RISC-V-based platforms: the Sophgo Sophon SG2042 System-on-Chip (SoC) \cite{sg2042}, the first RISC-V processor specifically designed for server applications. The SG2042 features a 64-core RISC-V CPU based on the Xuantie C920 architecture. Each core includes a 128-bit wide vector unit supporting RVV 0.7.1 for vector execution. The SoC provides 64 KB of L1 instruction and data caches per core, a 1 MB L2 cache shared among four-core clusters, and a 64 MB system-wide L3 cache. It supports four channels of 3200MHz ECC DDR4 memory and provides 32 PCIe Gen4 lanes. 
The Sophgo processor has been made commercially available for software development and prototyping with the Pioneer Milk-V board\cite{milkv}.
%The primary focus was on assessing the maturity of RISC-V HPC system software, addressing challenges in integrating multi-node RISC-V clusters, including developing an HPC stack with interconnects, storage, and power monitoring, all built on RISC-V hardware. 
%In this paper, we propose an upgrade of Monte Cimone named MCv2 to assess the maturity of software libraries and compilers in leveraging novel, more compute-capable RISC-V architectures featuring vector extensions and larger memory capacity.
%To address the performance limitations observed in MCv1, 
%We propose an upgrade of Monte Cimone named MCv2.%, which aims to demonstrate the feasibility of RISC-V for HPC-grade applications by enhancing the system's capabilities.
%The key contributions include: (i) integrating new RISC-V-based hardware with vector extensions, (ii) setting up an updated software stack with compilation toolchains and optimized BLAS libraries both from the community and produced in this work, and (iii) conducting performance benchmarking using HPC-class tools.

\section{Methods}
\subsection{Monte Cimone v2 hardware and software setup}
\begin{figure}[h!]
    \centering
    \includegraphics[width=0.8\columnwidth]{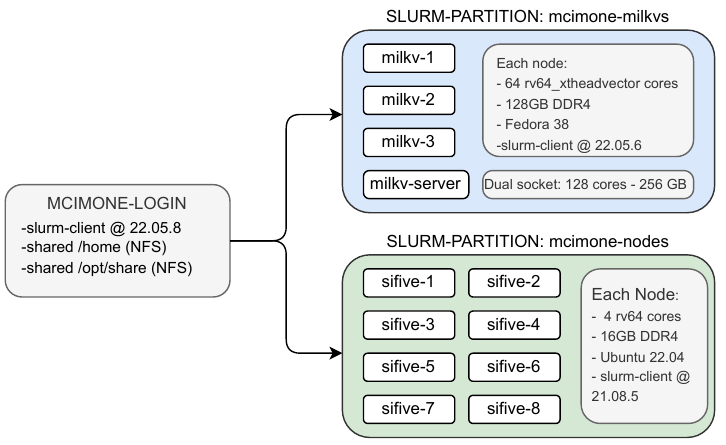}
    \caption{Monte Cimone  v1 (green) + v2 (blue) view}
    \label{fig:mcv1-v2}
\end{figure}
The first enhancement we introduced to the MCv2 setup is a hardware scale-out, expanding the original configuration with four additional nodes, each powered by the Sophgo Sophon SG2042. %%System-on-Chip (SoC)\footnote{SG2042 TRM, \url{github.com/milkv-pioneer/pioneer-files/blob/main/hardware/SG2042-TRM.pdf}.}, the first RISC-V processor designed for server applications. The SG2042 features a 64-core RISC-V CPU based on the Xuantie C920 architecture. Each core includes a 128-bit wide vector unit supporting RVV 0.7.1 for vector execution. The SoC provides 64 KB of L1 instruction and data caches per core, a 1 MB L2 cache shared among four-core clusters, and a 64 MB system-wide L3 cache. It supports four channels of 3200MHz ECC DDR4 memory and provides 32 PCIe Gen4 lanes.
Three of the nodes are Milk-V Pioneer Box systems, each featuring an SG2042 processor and 128 GB of memory. The fourth node is a dual-socket system with two SG2042 processors, providing a total of 128 cores and 256 GB of system memory, built on the Sophgo SR1-2208A0 platform.
The SG2042-based blades are integrated into the Monte Cimone system (i) using the existing 1Gb/s network and (ii) as an additional SLURM partition. Similarly, the MCv2 nodes have been configured using Spack and integrated into the ExaMon monitoring infrastructure.
%From the software perspective, 

The MCv2 nodes run Fedora 38 as the operating system, along with the upstream GCC 13 toolchain. To enhance compatibility with the Xuantie C920 core and its vector unit, we built and made available as shared modules two additional toolchains. The first one is the Xuantie GNU Toolchain \cite{xuantie-gcc}, a customized GCC 10-based toolchain specifically designed by the Xuantie core developer for compiling code targeting the RVV 0.7.1 vector extension. The second one is the GNU GCC 14 toolchain, which introduces support for the \texttt{theadvector} compilation target, as GNU GCC identifies the vector extension of the C920 core. The overall structure of the system can be seen in Figure \ref{fig:mcv1-v2}.

\subsection{MCv2 performance analysis tools}
To assess MCv2, we carried out a series of tests focusing on memory performance and FP64 scalar and vector execution. For these evaluations, we employed the STREAM benchmark to measure memory bandwidth and the HPL benchmark to assess high-performance computing capabilities.
STREAM and HPL were compiled with GCC 13 with the latter linked against two different sets of OpenBLAS libraries \cite{openblas}. The first configuration used OpenBLAS built for the generic RV64 target, serving as a baseline that does not leverage the processor’s vector unit. 
The second configuration utilized an optimized version of OpenBLAS, incorporating assembly kernels specifically designed for the C920 core and its vector unit. These optimized kernels, available in the official OpenBLAS repository, were compiled with the appropriate architectural target using the Xuantie GNU Toolchain. 

\subsection{MCv2 software stack enhancement: BLIS optimization}
In addition to the initial setup and characterization of MCv2, our efforts were dedicated to expanding its software stack. A key focus of our work was integrating an alternative set of BLAS libraries known as BLIS \cite{blis}.
BLIS is an open-source BLAS implementation designed to enhance portability across emerging microarchitectures, offering an alternative to other open-source libraries like OpenBLAS. Its framework contains small computational units called micro-kernels that, wrapped in different ways called macro-kernels, implement different BLAS functions.  This blocking is exposed to the programmer facilitating efficient cache utilization and micro-kernel optimization. In our work, we leverage these features to develop a viable alternative to OpenBLAS for MCv2 and, more broadly, for the SG2042 processor.

\subsubsection{Retrofitting from RVV 1.0 to RVV 0.7.1}

By default, BLIS includes assembly micro-kernels written for RVV 1.0, enabled via the \texttt{rv64iv} target. However, by adapting the microkernels to RVV 0.7.1, following our translation process, these libraries can be built for the \texttt{theadvector} machine architecture supported by GCC 14. 
This process involved adapting \textit{load} and \textit{store} instructions, as well as \textit{vsetvl} operations to the older syntax. Additionally, a \textit{th.} prefix was added to each vector instruction, enabling the compiler to recognize them as \texttt{theadvector}. %Once the adapted version of BLIS was compiled, we could rerun the previous tests using HPL to evaluate the performance of the provided RVV kernels. %From Figure \ref{fig:blis}, specifically the first and second columns of each group, we can observe the absolute performance of BLIS with the provided vector kernels compared to the optimized OpenBLAS. These results serve as the starting point for our optimization process, aiming to develop another BLAS library that achieves at least the same level of performance.
The adapted and compiled micro-kernels were evaluated using the same HPL benchmark, with results obtained from OpenBLAS serving as the baseline. As discussed in Section \ref{subsec:blis_ev}, there was still room for improvement, prompting the start of the optimization process.
\subsubsection{BLIS optimization}
\label{subsec:blis_opt}
The first step in our optimization process was to assess the primary bottleneck, which could stem from either inefficient cache utilization or suboptimal micro-kernel code. As discussed in Section \ref{subsec:blis_ev} the main performance bottleneck of BLIS is the second.
\begin{figure}[h]
    \centering
    \begin{subfigure}{0.48\textwidth}
        \centering
        \includegraphics[width=\textwidth]{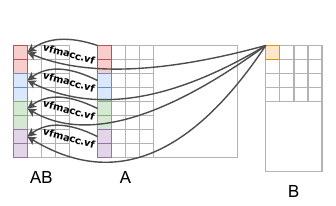}
        \caption{Original BLIS micro-kernel}
        \label{fig:sub1}
    \end{subfigure}
    \hfill
    \begin{subfigure}{0.50\textwidth}
        \centering
        \includegraphics[width=\textwidth]{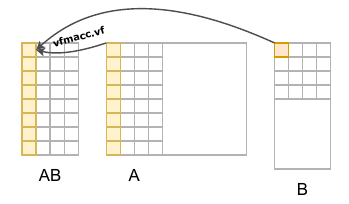}
        \caption{Optimized BLIS micro-kernel}
        \label{fig:sub2}
    \end{subfigure}
    \caption{Focus point of our micro-kernel optimization}
    \label{fig:main}
\end{figure}

The micro-kernel targeted for optimization serves as the foundation for level-3 BLAS functions, such as GEMM, and is primarily composed of rank-1 updates. The original implementation, illustrated in Figure \ref{fig:sub1}, operates on single vector registers, repeatedly invoking the \textit{vfmacc.vf} instruction on contiguous data. This design choice was likely intended to maintain microarchitecture agnosticism and ensure a reusable RVV-based micro-kernel.\\
We optimized this implementation by reducing the number of fetched instructions while preserving the existing data blocking and algorithm. On the SG2042 processor, each vector register holds two FP64 values. Consequently, updating an eight-element column of
AB requires four \textit{vfmacc.vf} calls and four load operations to populate four vector registers with a column of A.
To enhance efficiency, we leveraged register grouping by increasing the RVV LMUL parameter from one to four, with a subsequent remap of data across vector registers. This adjustment allows a single load operation to populate four vector registers with an entire column of A, and a single \textit{vfmacc.vf} instruction to update a column of AB, as illustrated in Figure \ref{fig:sub2}.

\section{Experimental Results}
In this section, we summarize the achieved performance in memory bandwidth using the STREAM benchmark and in FP64 execution using HPL. The latter is analyzed from two perspectives. First, we present a system-wide characterization of MCv2 using HPL linked to OpenBLAS libraries. Second, we evaluate the impact of our BLIS porting and optimization efforts.

%To evaluate MCv2, we carried out a series of tests similar to those conducted on MCv1, focusing on memory performance and FP64 scalar and vector execution. These tests utilized the STREAM and HPL benchmarks. STREAM and HPL were compiled with GCC 13 with the latter linked against two different sets of OpenBLAS\footnote{OpenBLAS repository, \url{github.com/OpenMathLib/OpenBLAS}.} libraries. The first configuration used OpenBLAS built for the generic RV64 target, serving as a baseline that does not leverage the processor’s vector unit. The second configuration utilized an optimized version of OpenBLAS, incorporating assembly kernels specifically designed for the C920 core and its vector unit. These optimized kernels, available in the official OpenBLAS repository, were compiled with the appropriate architectural target using the Xuantie GNU Toolchain. %The second one is BLIS\footnote{BLIS repository, \url{github.com/flame/blis}.}, that provides assembly kernels intended for RVV 1.0, adapted in order to be compiled by the GCC 14 toolchain with \textit{theadvector} target. 
\subsection{MCv2 Memory Performance} %Figure \ref{fig:stream_64omp} reports the collected STREAM results. 
\begin{figure}[h!]
     \centering
     \includegraphics[width=0.85\columnwidth]{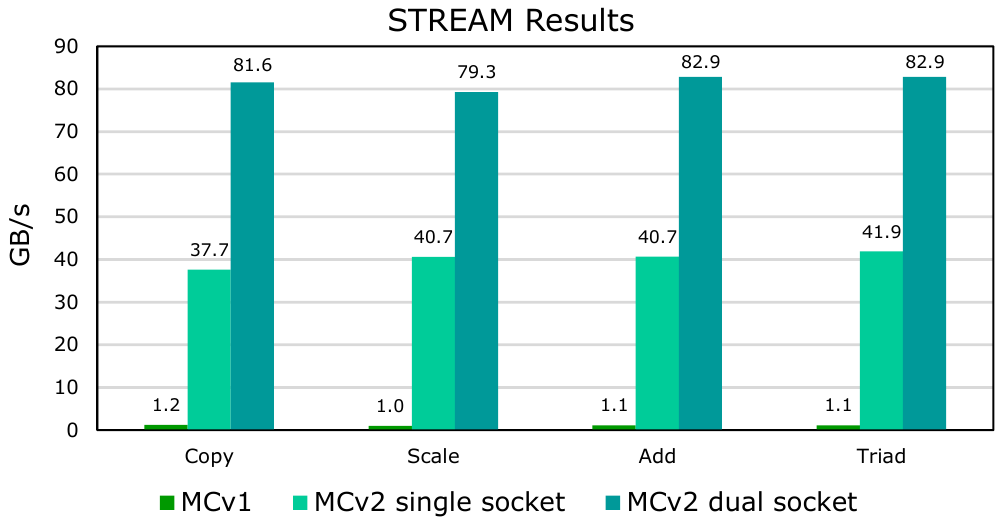}
     \caption{STREAM benchmark on a MCv2 node with 64 OpenMP threads compared to a MCv1 node}
     \label{fig:stream_64omp}
 \end{figure}

%This paragraph describes the STREAM results.
The MCv2 single socket node saturates its memory bandwidth with 64 OpenMP threads, achieving a bandwidth of 41.9 GB/s. Interestingly, the  MCv2 dual socket node achieves a memory bandwidth of 82.9 GB/s, still using 64 OpenMP threads but pinned symmetrically in the two sockets - increasing the number of OpenMP threads reduces the attained bandwidth. In contrast, an MCv1 node achieves a memory bandwidth of 1.1 GB/s with 4 OpenMP threads. A comprehensive view of this data is visible in Figure \ref{fig:stream_64omp}.
%However, we observed nonlinear memory performance scaling with lower thread count runs. This helps to explain the lack of linear scaling in HPL benchmarks performed at varying core counts on a single node.
%\paragraph{HPL}
\subsection{MCv2 FP64 Performance}
\begin{figure}[h!]
    \centering
    \includegraphics[width=\columnwidth]{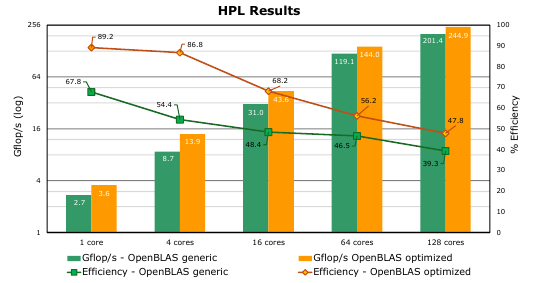}
    \caption{MCv2 HPL w. OpenBLAS (generic \& optimized compiling target)}
    \label{fig:hpl_1node}
\end{figure}

Figure \ref{fig:hpl_1node} reports the performance characterization for the HPL benchmark for the MCv2 compute node while scaling the number of cores and for different BLAS libraries. From the Figure, we can see that the vanilla OpenBLAS libraries are still lagging behind the SG2042-optimized ones, with a relative efficiency of 68\% with one core, which increases to 89\% of the optimized one. Both of them experience a degradation in relative performance when all the cores are used. This suggests that the optimized OpenBLAS suffers from SoC's bottlenecks as the unoptimized one.

%As shown in Figure \ref{fig:hpl_1node}, performance scaling is not strictly linear, and efficiency declines as core count increases. At lower core counts, efficiency, measured as the achieved percentage of the theoretical peak, approaches 90\%. However, as the number of cores increases, the performance gain diminishes.
%In summary, the achievable performance per core is approximately 3.5 Gflop/s, scaling up to nearly 144 Gflop/s across 64 cores. 
%These results demonstrate a significant performance improvement compared to the previous SiFive hardware platform, aligning with already available comparison between the SiFive Freedom U740 SoC and the Sophgo Sophon SG2042\cite{nickbrown2025}, though obtained using a different software stack. Additionally, our work extends this analysis to a multi-node cluster with also a dual socket platform, providing further insights into performance scaling across multiple MCv2 nodes. 
These results confirm previous work comparison between the SiFive Freedom U740 SoC and the Sophgo Sophon SG2042 \cite{nickbrown2025}, though obtained using a different software stack. Furthermore, we extend previous works to a multi-node cluster and dual-socket nodes, providing further insights into performance scaling across multiple MCv2 nodes. 

\begin{figure}[h!]
    \centering
    \includegraphics[width=0.8\columnwidth]{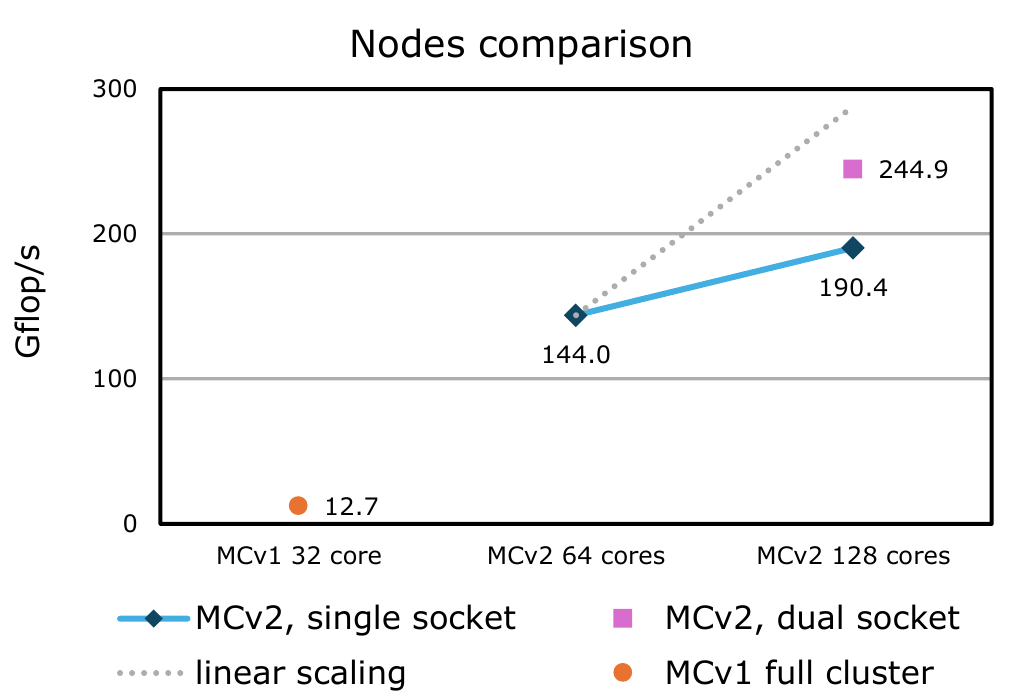}
    \caption{HPL on different node's configurations}
    \label{fig:hpl_multi_node}
\end{figure}

 Figure \ref{fig:hpl_multi_node} shows the performance results from HPL runs with different node configurations. The MCv1 32-cores case refers to the HPL executed in parallel on all 8 MCv1 compute nodes, which achieves 13 Gflop/s. The MCv2 64-cores case refers to the MCv2 single-socket HPL run, while the MCv2 128-cores case refers to both (i) a dual MCv2 single socket nodes configuration and (ii) a single MCv2 dual socket node configuration. From it, we can notice that while in MCv1, the 1 Gb/s network was sufficient for obtaining almost an HPL linear scaling, in the case of the performance of the MCv2 nodes, it is no longer sufficient and increasing the number of parallel processes reduces the HPL efficiency (only the 1.33$\times$ w.r.t single node performance). The MCv2 dual-socket compute node, in contrast, achieves almost 1.76$\times$ of the performance of the MCv2 single-socket node and 127$\times$ more performance of an MCv1 compute node. 

\subsection{BLIS porting and optimization evaluation}
\label{subsec:blis_ev}
\begin{figure}[h!]
    \centering
    \includegraphics[width=0.9\columnwidth]{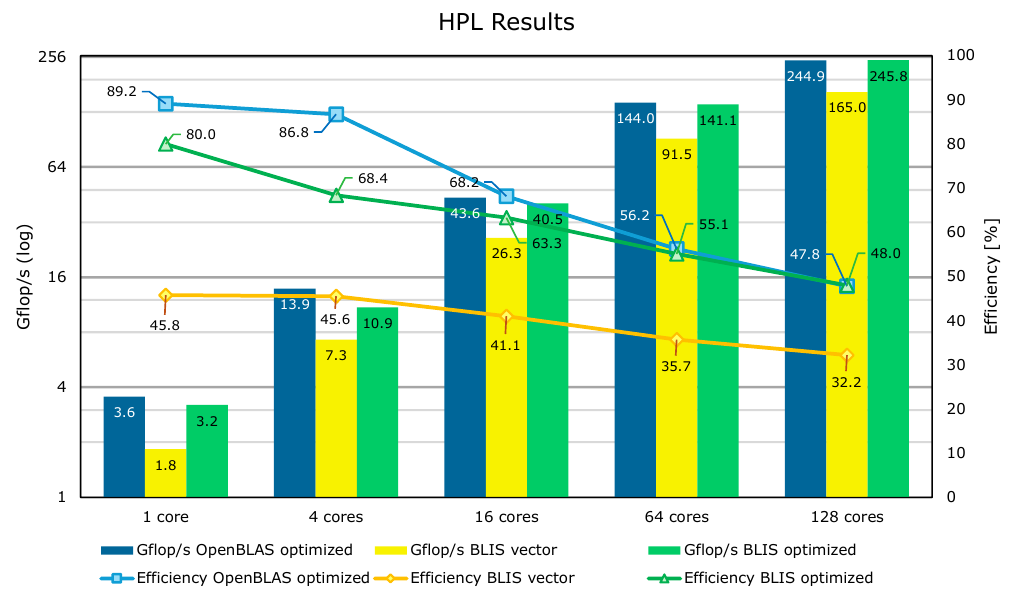}
    \caption{MCv2 attained performance comparison between HPL+OpenBLAS vs HPL+BLIS pre/post-optimization.}
    \label{fig:blis}
\end{figure}

With a performance baseline for MCv2 established using HPL and OpenBLAS, we can now evaluate the impact of our work on the BLIS library.
 %From Figure \ref{fig:blis}, specifically the first and second columns of each group, we can observe the absolute performance of BLIS with the provided vector kernels compared to the optimized OpenBLAS.
Figure \ref{fig:blis} provides a comprehensive view of HPL benchmark results across different libraries and core counts. The first column for each core count represents our baseline, obtained from HPL linked to the optimized OpenBLAS, while the second column illustrates the performance of HPL using BLIS with the provided vector micro-kernels. The third column of each core count group contains the final results of HPL linked to our optimized version of BLIS. The first two served as the foundation for our optimization process, with the goal of developing a BLIS version that achieves performance on par with, or exceeding, that of OpenBLAS.
After obtaining the initial performance results for BLIS, our first objective was to identify and address the primary performance bottleneck. Since BLIS optimization primarily follows two approaches—cache blocking adjustments and micro-kernel optimization—we conducted a comparative analysis of cache misses to determine the most effective path forward. This analysis involved running HPL and measuring cache miss data using Linux \texttt{perf} \cite{perf}.
\begin{figure}[h!]
    \centering
    \includegraphics[width=0.9\columnwidth]{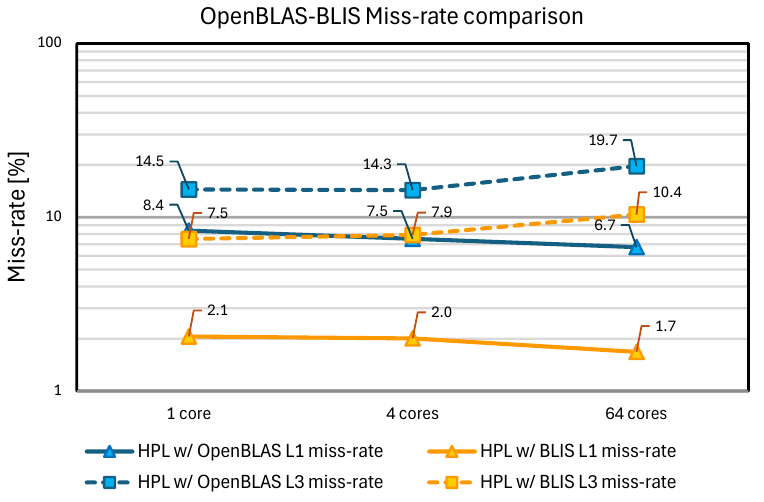}
    \caption{MCv2 cache Miss-rate: HPL+OpenBLAS vs HPL+BLIS.}
    \label{fig:missrate}
\end{figure}
Figure \ref{fig:missrate} presents the results that guided our decision to focus on micro-kernel refinement. The horizontal axis represents various core counts, while the vertical axis shows the cache miss rate for L1 and L3 caches during HPL runs linked to both optimized OpenBLAS and non-optimized BLIS. The data clearly indicate that non-optimized BLIS already exhibits superior cache performance (L1-to-L1 and L3-to-L3 comparison) compared to optimized OpenBLAS. This observation led us to conclude that the FP64 performance of BLIS is hindered by inefficiencies in the provided vector micro-kernels.
%As shown in Figure \ref{fig:missrate}, HPL linked against BLIS libraries, even without optimizations, already outperforms the same benchmark linked against OpenBLAS in terms of cache performance. This observation led us to conclude that the bottleneck lies in the provided microkernel code, prompting us to focus on its optimization.
With the reduction in the instruction count of the micro-kernel described in Section \ref{subsec:blis_opt}, we conducted tests to evaluate its impact. As shown by the third column of each core count group in Figure \ref{fig:blis}, there is a significant improvement in attainable performance compared to the original BLIS micro-kernel. The results are now comparable to those of OpenBLAS and, in some cases, even superior. For example, the most noticeable case occurs when running HPL on 128 cores. In these tests, the baseline performance with HPL linked to optimized OpenBLAS is 244.9 Gflop/s, while the non-optimized BLIS implementation achieves only 165.0 Gflop/s. Through our optimizations, BLIS surpasses OpenBLAS, reaching 245.8 Gflop/s, representing a 49\% improvement over the baseline BLIS implementation.

%We measured a computing performance improvements of 127x coupled with a memory bw increase of 69x. The HPL performance of a dual-socket RV node of MV v2 is is now comparable to an HPC server node in 20219??  while MC v1 node would be comparable to an HPC node in 2009?  A 10 years progress has been achieved in just 2 years., indicating that RV is rapidly becoming a viable architecture for future supercomputers.
\section{Conclusions}
This work presents a comparative analysis of MCv1 and MCv2 using standard HPC evaluation tools. Our findings highlight the rapid evolution of the RISC-V ecosystem within just two years. For context, the Top500 list reports an average 127× in performance improvement over eight years \cite{top500}, underscoring the accelerated development of RISC-V hardware as a viable platform for HPC systems. Furthermore, our efforts in porting and optimizing BLAS libraries, specifically BLIS, demonstrate the feasibility of enhancing the RISC-V software ecosystem, ultimately benefiting the broader HPC community.

\small
\subsubsection*{Acknowledgements} This activity has been supported by the HE EU Graph-Massivizer (g.a. 101093202), DECICE (g.a. 101092582), and DARE (g.a. 101143421) projects, as well as the Italian Research Center on High Performance Computing, Big Data, and Quantum Computing.
\normalsize

\end{document}